\documentclass[%
 reprint,
 amsmath,amssymb,
 aps,
pra,superscriptaddress,
longbibliography
]{revtex4-2}

\usepackage{graphicx}
\usepackage{dcolumn}
\usepackage{bm}
\usepackage{color}
\usepackage{hyperref}

\begin{document}

\title{Restricted-Geometry Quantum Models Beyond Atoms:\\ Application of the Eckhardt-Sacha approach to NSDI in Diatomic Systems}

\author{Lars C. Bannow}%
\affiliation{%
 Fachbereich Physik, Philipps-Universit\"at Marburg, D-35032 Marburg, Germany
}%

\author{Jan H. Thiede}%
\affiliation{%
  Fachbereich Physik, Philipps-Universit\"at Marburg, D-35032 Marburg, Germany
}%

\author{Michał Ogryzek}
\affiliation{Institute of Theoretical Physics, Faculty of Fundamental Problems of Technology,
Wrocław University of Science and Technology, 50-370 Wrocław, Poland}

\author{Dmitry K. Efimov}
\affiliation{Institute of Theoretical Physics, Faculty of Fundamental Problems of Technology,
Wrocław University of Science and Technology, 50-370 Wrocław, Poland}

\author{Jakub S. Prauzner-Bechcicki}%
\email{jakub.prauzner-bechcicki@uj.edu.pl}
\affiliation{%
 Jagiellonian University in Kraków, Faculty of Physics, Astronomy and Applied Computer Science,
Marian Smoluchowski Institute of Physics, Łojasiewicza 11, 30-348 Krakow, Poland
}%

\date{\today}

\begin{abstract}
We present a (1+1)-dimensional quantum model designed to describe nonsequential double ionization (NSDI) in homonuclear diatomic molecules exposed to strong linearly polarized laser fields. Extending the restricted-geometry framework previously developed for atomic systems by Eckhardt and Sacha, our approach captures key features of NSDI, including the characteristic knee structure in double ionization yields. Despite its simplifying assumptions, the model shows good agreement with experimental data and proves particularly suitable for systems with $\sigma$-type orbital symmetry. It offers a computationally efficient tool for exploring multi-electron dynamics in molecular systems.
\end{abstract}


\maketitle

\section{\label{sec:intro}Introduction}
Experimental observations of strong-field phenomena such as above-threshold ionization (ATI), high-harmonic generation (HHG), and non-sequential ionization (NSI) have posed significant challenges for theoretical interpretation. While acknowledging the many efforts and diverse approaches, it seems justified to state that breakthroughs in our understanding of these processes have often come from simple models that offered intuitive explanations and indicated where new phenomena might be expected. A prominent example is the single active electron (SAE) approach, followed by the strong-field approximation (SFA), which significantly advanced in particular the theoretical description of HHG, NSI, and other strong-field phenomena~\cite{Lewenstein1994,Amini2019}. Interestingly, this highly successful method was preceded by an even simpler classical concept, the so-called simple-man model, also known as the rescattering scenario~\cite{Corkum93,Kulander1993}.

These developments were accompanied by the extensive use of another approach that greatly simplifies calculations and provides intuitive explanations of the studied phenomena~\cite{Javanainen1988,Eberly1989,Eberly1990,Su1990,Pindzola1991,Grobe1992,Grobe93,Liu1999,Becker12}.
This approach is based on the observation that, in a linearly polarized field, the dominant influence on electron dynamics occurs along the polarization axis.
Consequently, the generally three-dimensional problem for a single electron can be reduced to a one-dimensional model that captures the essential physics of the analyzed effect.
For non-sequential double ionization (NSDI), this simplification results in a $(1+1)$-dimensional model, with one coordinate along the laser polarization axis for each electron~\cite{Grobe1992,Grobe93,Liu1999,Lein00,Becker12}.

Building on these simplified and restricted-geometry approaches, Eckhardt and Sacha introduced a new perspective on NSDI of atoms~\cite{Eckhardt2001physicaScripta,Eckhardt01,Sacha01,Sacha2001triple,Sacha2003,Sacha06}, which culminated in the development of a novel restricted-geometry quantum model~\cite{Prauzner-Bechcicki2007,Prauzner-Bechcicki2008}. The key idea of this approach is to constrain the escaping electrons to evolve within a symmetric subspace that naturally reflects the correlated two-electron motion characteristic of NSDI, thereby isolating the physically relevant degrees of freedom while retaining the essential recollision-driven dynamics.
This model successfully reproduced and explained key experimental observations, such as the knee structure in the ionization yield curve~\cite{LHuillier82,LHuillier83a,Walker94,Larochelle98}, the double-hump structure in the ion momentum distribution~\cite{Weber00,Moshammer00}, and characteristic structures in two-electron momentum distributions~\cite{Correlation00,Staudte07,Rudenko07}, while enabling systematic analysis of NSDI across a broad range of laser-field parameters~\cite{Prauzner-Bechcicki20072,Eckhardt2008,Eckhardt20082,Eckhardt2010,Jiang2023,Thiede2024,Liu2025}.

The approach was subsequently compared to other restricted-geometry concepts~\cite{Efimov2018} and later extended to describe triple ionization~\cite{Thiede2018,Efimov2018,Efimov2019,Efimov2020,Prauzner-Bechcicki2021,Efimov2021,Efimov2023,Efimov2025}. In this work, we further advance the restricted-geometry quantum framework, originally proposed by Eckhardt and Sacha, to explore its predictive power and limitations in describing multi-electron dynamics under strong laser fields. Particular emphasis is placed on the mechanisms of non-sequential and sequential ionization processes in diatomic molecules.

Available data for diatomic and other small molecules clearly exhibit signatures of non-sequential double ionization (NSDI) similar to those observed in atoms, while also 
revealing important differences~\cite{Talebpour97,Cornaggia98,Guo98,Guo00,Alnaser2003-tx,Alnaser2004-rm,Eremina04,Zeidler05,Zeidler2006,Chen2007-fa,Herrwerth2008,Baier2008-xj,Liao:09,Jia09,Mauger2009-zz,Wu2010,Huang2011-xp,Dura2012,Lin2012,Emmanouilidou2012-rj,Kubel2013,Jia2013-vu,Li2014-ix,Wang2015-jy,Cheng2019,Li2019-xg,Li2020-fx,Hanus2020-rq,Howard2021,Cheng2021,Ren2022-gk,Yuen2022-rn,Yuen2024-xo,Jia2024-pv}.
These differences stem from the richer internal structure of molecules compared to atoms and the presence of additional degrees of freedom, such as the existence of a molecular axis.
They manifest as suppressed ionization yields, modified momentum distributions, and angular dependencies; for example, sensitivity to the alignment between the molecular axis and the laser polarization direction.

The double ionization of diatomic molecules has been actively studied over the past 25 years. Experimentally, the double ionization yield has been investigated in H$_2$ and D$_2$ as a function of laser field polarization~\cite{Alnaser2003-tx}, and in O$_2$ and N$_2$ with respect to orbital configuration~\cite{Alnaser2004-rm}. Studies on heteronuclear diatomic molecules have focused on the role of multiorbital contributions~\cite{Li2019-xg}, and were later extended to polyatomic systems in linearly polarized fields, with particular attention to NSDI efficiency~\cite{Li2020-fx}. Other experiments have targeted subcycle resolution of H$_2$ ionization~\cite{Hanus2020-rq} and the analysis of momentum distributions following double ionization of homonuclear molecules~\cite{Zeidler05,Ren2022-gk}.

In parallel, a wide range of theoretical and numerical methods have been developed to interpret these findings. Classical trajectory simulations have been extensively used to study the double ionization knee~\cite{Liao:09,Li2014-ix}, with some approaches incorporating tunneling of the first electron via ADK-based initial conditions~\cite{Chen2007-fa,Emmanouilidou2012-rj}, while others relied on purely classical dynamics~\cite{Mauger2009-zz,Huang2011-xp}. Analytical S-matrix methods have also been applied to explore alignment-dependent effects~\cite{Jia09,Jia2013-vu}.

Quantum models, including one-dimensional and cylindrical (3D for two electrons) approaches, have enabled the study of ionization channels and alignment effects~\cite{Baier2008-xj}. The introduction of a four-dimensional model (two spatial dimensions per electron)~\cite{Wang2015-jy} allowed for the investigation of two-center interference in molecular double ionization. More recently, a density-matrix-based semi-analytical framework has been proposed to describe sequential double ionization in molecules~\cite{Yuen2022-rn,Yuen2024-xo,Jia2024-pv}.

It is evident that models developed for atomic systems fall short in capturing molecular-specific effects. To address this, theoretical approaches must be adapted or extended to incorporate parameters such as internuclear separation and orientation-dependent ionization pathways, just to mention the simplest ones. Yet, moving from atomic to molecular models inevitably increases computational complexity and the numerical cost of full-dimensional simulations. This raises a fundamental and still open question: to what extent can simplified restricted-geometry models be reliably applied to molecular systems?
Before introducing our model, we note that simplified approaches to molecular NSDI also necessarily rely on assumptions such as fixed nuclei and the use of smoothed Coulomb interactions; these aspects, together with their implications, are discussed in Sec.~\ref{model}.

As long as one considers models in which linear polarization of the laser field defines a privileged direction of electron motion, the geometric reduction is straightforward and follows the same logic as in the atomic case. However, when the restricted-geometry model originally derived by Eckhardt and Sacha for atoms~\cite{Eckhardt2001physicaScripta,Eckhardt01,Sacha01,Sacha2001triple,Sacha2003,Sacha06} is to be applied to molecular systems, its validity and implementation require careful justification. A classical extension of this approach to molecules was proposed in~\cite{Prauzner-Bechcicki2005,Prauzner-Bechcicki20052}, but a quantum counterpart has not been systematically developed or evaluated.

In this work, we extend the restricted-geometry quantum framework within the Eckhardt–Sacha approach to diatomic molecules and assess its ability to capture both universal and molecule-specific features of strong-field ionization. Our findings aim to clarify the applicability limits of this method and to identify which molecular NSDI mechanisms can be reproduced within such reduced-dimensionality models. Additionally, the approach offers a computationally efficient tool for studying multi-electron dynamics in systems where full-dimensional quantum calculations remain prohibitively expensive.

The paper is organized as follows. In Section II, we discuss the rationale behind the restriction of geometry. Section III briefly outlines the numerical approach. In Section IV, we present and analyze the ionization yields and momentum distributions. Finally, conclusions are drawn in Section V. Atomic units are used throughout the paper unless stated otherwise.

\section{Rationale for Restricted Geometry\label{model}}
We consider atoms or molecules interacting with a strong, short, linearly polarized laser pulse. The $z$-axis is the polarization axis. Below, we provide a concise overview of the approach that enables dimensionality reduction in the case of atoms.  A more detailed discussion can be found in previous  papers~\cite{Eckhardt2001physicaScripta,Eckhardt01,Sacha01,Sacha2001triple,Sacha2003,Sacha06,Prauzner-Bechcicki2007,Prauzner-Bechcicki2008}.

Nowadays, the rescattering scenario~\cite{Corkum93,Kulander1993} is widely accepted as the underlying mechanism of NSDI. In this three-step process, electrons assist each other in escaping the atomic or molecular potential. In the first step, one electron escapes near a field maximum via tunneling or over-the-barrier ionization. It is then accelerated by the laser field and driven back towards the core when the field reverses its sign. Upon recollision, the second electron may be released either immediately (so-called direct ionization) or after a delay, following excitation by the first electron and subsequent field ionization (recollision-induced excitation with subsequent ionization, RESI). Other mechanisms are also possible, such as the slingshot scenario~\cite{Katsoulis2018} or multiple recollision cycles.

The last step of the rescattering scenario can also be viewed from a slightly different, adiabatic perspective. Upon interaction with the ion and the second electron remaining near the core, the returning electron may become “captured,” forming a doubly excited complex~\cite{Eckhardt2010,Camus12}. The quotation marks emphasize that this complex is typically interpreted either as a virtual intermediate state or as a real, short-lived transient Coulomb complex. In this picture, the observed effects of rescattering are understood as decay channels of this transient state, leading to various outcomes: simultaneous ionization of both electrons (direct ionization), delayed ionization (RESI), or even purely sequential double ionization.  

Decay channels of the transient state can be identified through a local stability analysis of the adiabatic full-dimensional potential~\cite{arnold2013mathematical}. This approach has been successfully applied to the study of NSDI in atoms~\cite{Eckhardt2001physicaScripta,Eckhardt01,Sacha01,Sacha2001triple,Sacha2003,Sacha06}, leading to a key observation: simultaneous escape of both electrons, responsible for signatures of correlated motion visible in ion momentum distributions and two-electron momentum maps, is possible only in the vicinity of a specific subspace. In this subspace, the electrons move symmetrically with respect to the laser polarization axis and pass sufficiently close to saddle points formed by the interplay between Coulomb interactions and the external electric field. These saddles act as gateways for correlated ionization, and their geometry and accessibility strongly influence the likelihood of direct double ionization.

In the case of atoms, the analysis of saddle points has shown that their position depends on both the field amplitude and the number of active electrons considered. While the former determines the distance between the saddle and the nucleus, the latter constrains the motion of the saddle in configuration space. When visualized in three-dimensional space, the saddle lies on a straight line inclined at a constant angle of $\alpha = \pm \pi/6$ with respect to the laser polarization axis, at a distance from the nucleus $r_s$  with (see ~\cite{Sacha01})
\begin{equation}
    r_s^2 = \frac{\sqrt{3}}{|F(t)|},
\end{equation}
where $F(t)$ is the instantaneous electric field amplitude at time $t$. Thus, the position of the saddle in cylindrical coordinates is given by (see ~\cite{Sacha01})
\begin{equation}
    \left\{ \begin{array}{cc}
         Z_S=r_s\cos \alpha&  \\
         R_S=r_s\sin \alpha& 
    \end{array}
    \right. ,
    \label{lines}
\end{equation}
as illustrated in Fig. 1 of that reference.
Since the saddle acts as a gateway for the correlated escape of electrons, their motion can be effectively confined to two lines defined by Eq.~(\ref{lines}), resulting in a (1+1)-dimensional model atom~\cite{Prauzner-Bechcicki2007,Prauzner-Bechcicki2008}.

A first extension of the above approach to diatomic homonuclear molecules was presented in~\cite{Prauzner-Bechcicki2005,Prauzner-Bechcicki20052}, where the saddles of the adiabatic full-dimensional potential were analyzed, the symmetric $C_{2v}$ subspaces were identified, and classical simulations of the decay process were performed, yielding ion momentum and two-electron momentum distributions for N$_2$ and O$_2$ molecules. In the present work, we extend this analysis by developing a fully quantum model. But before proceeding, we briefly recap the structure of the identified subspaces and the properties of the corresponding saddle points.

\subsection{Symmetric subspaces and saddle structure}

Since the proposed model for diatomic homonuclear molecules is an extension of the approach successfully applied to atomic systems, we will use the atomic case (referred to as the helium atom) as a benchmark. Throughout this analysis, we neglect nuclear motion by assuming that the nuclei are infinitely heavy compared to the electrons. The starting point is a hydrogen-like Hamiltonian of the form
\begin{equation}
 H = \frac{\mathbf{p}_1^2}{2}+\frac{\mathbf{p}_2^2}{2}+V,
\end{equation}
where $\mathbf{p}_i$ denotes the momentum of the $i$-th electron, and $V$ is the total potential consisting of two parts:
\begin{equation}
    V = V_C+V_I,
    \label{pot}
\end{equation}
here, we denote the Coulomb term by $V_C$ and the field interaction term by $V_I$, respectively.
The field interaction term is similar for both helium and the molecule. Using the dipole approximation and assuming a linearly polarized laser field $F(t)$ along the $z$-axis, it takes the form
\begin{equation}
    V_I=(z_1+z_2)F(t).
    \label{interacion}
\end{equation}
The Coulomb term, on the other hand, distinguishes helium from molecular systems. Specifically, for helium it reads
\begin{equation}
    V_C^{He} = -\sum_{i=1}^2\frac{2}{r_i} + \frac{1}{|\mathbf{r}_1-\mathbf{r}_2|},
    \label{He_pot}
\end{equation}
whereas for a diatomic molecule it is given by
\begin{equation}
 \begin{split}
  V_C^{M} & = -\sum_{i=1}^{2}\frac{1}{\sqrt{\left(x_i+\frac{d}{2}\sin{\theta}\right)^2+y_i^2+\left(z_i+\frac{d}{2}\cos{\theta}\right)^2}}\\
      &-\sum_{i=1}^{2}\frac{1}{\sqrt{\left(x_i-\frac{d}{2}\sin{\theta}\right)^2+y_i^2+\left(z_i-\frac{d}{2}\cos{\theta}\right)^2}}\\
      &+\frac{1}{|\mathbf{r}_1-\mathbf{r}_2|}. 
 \end{split}
 \label{Mol_pot}
\end{equation}
Here, the center of mass of the nuclei is placed at the origin of the coordinate system and $d$ is the distance between the nuclei. Without loss of generality, the nuclei are assumed to lie in the $xz$--plane, and $\theta$ denotes the angle between the molecular axis and the laser polarization axis. In both cases, $\mathbf{r}_i=(x_i,y_i,z_i)$ denotes the position of the $i$-th electron.

\begin{figure}[tb!]
 \begin{centering}
  \includegraphics[scale=1.]{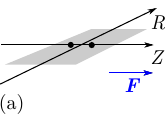}
  \includegraphics[scale=1.]{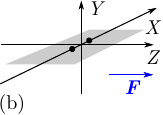}
  \includegraphics[scale=1.]{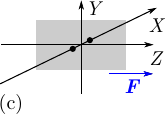}
  \caption{\label{fig:config}(Color online)
  The three configurations leading to the model potentials $V_{\parallel}$ (a), $V_{\perp 2}$ (b), and $V_{\perp 3}$ (c). Black dots represent the protons, the gray shaded plane indicates the plane in which the electrons move, and $\bm{F}$ denotes the field polarization axis, which is aligned with the $Z$-axis in all cases. In panel (c), one proton is located in front of the gray plane, and the other behind it.}
 \end{centering} 
\end{figure}

Local analysis of the adiabatic potential, Eq.~(\ref{pot}), with the molecular Coulomb term given by Eq.~(\ref{Mol_pot}), allowed Prauzner-Bechcicki \textit{et al.} to identify three two-dimensional, $C_{2v}$-symmetric subspaces (see~\cite{Prauzner-Bechcicki2005} for details). These configurations are illustrated in Fig.~\ref{fig:config}.

In the first configuration, the molecular axis is aligned with the field polarization axis, i.e., $\theta = 0$ (Fig.~\ref{fig:config}(a)). Since this is a \textit{parallel} alignment, we denote the corresponding potential by $V_{\parallel}$, which includes both Coulomb and interaction terms. Introducing cylindrical coordinates due to the axial symmetry of the configuration and restricting the electrons to symmetric motion in a plane (i.e., $r_i=R$, $\varphi_1 - \varphi_2 = \pi$, $z_i=Z$), we obtain
\begin{equation}
\begin{split}
V_{\parallel}(R,Z,t)& = -\frac{2}{\sqrt{R^2+(Z+d/2)^2}}\\
&-\frac{2}{\sqrt{R^2+(Z-d/2)^2}}+\frac{1}{2|R|}+2Z F(t).
\label{eq:potA}
\end{split}
\end{equation}

The second configuration corresponds to the case where the molecular axis lies in the $XZ$-plane and is perpendicular to the field polarization direction, i.e., $\theta = \pi/2$. The electrons are restricted to symmetric motion in the $XZ$-plane (Fig.~\ref{fig:config}(b)). The resulting potential is
\begin{equation}
\begin{split}
V_{\perp 2}(X,Z,&t) = -\frac{2}{\sqrt{Z^2+(X+d/2)^2}} \\
&-\frac{2}{\sqrt{Z^2+(X-d/2)^2}}+\frac{1}{2|X|}+2Z F(t). 
\end{split}
\label{eq:potB}
\end{equation}
We denote this configuration by $V_{\perp 2}$, as the molecular axis and the field polarization axis are \textit{perpendicular}, and the electrons move in the plane spanned by both axes. All components of the model are thus confined to a two-dimensional plane.

Finally, we consider the case where the molecular axis is again \textit{perpendicular} to the field polarization axis ($\theta = \pi/2$), but the electrons are restricted to symmetric motion in the $YZ$-plane, while the nuclei remain in the $XZ$-plane (Fig.~\ref{fig:config}(c)). Although the dynamics take place in the $YZ$-plane, 
the configuration is “quasi” \textit{three-dimensional}, and we denote the potential by 
\begin{equation}
    V_{\perp 3}(Y,Z,t) = -\frac{4}{\sqrt{Y^2+Z^2+d^2/4}}+\frac{1}{2|Y|}+2Z F(t).\label{eq:potC}
\end{equation}

\subsection{Restricted-Geometry Hamiltonian}

Sacha and Eckhardt suggested in~\cite{Sacha01,Eckhardt01} that correlated double ionization requires the electrons to escape simultaneously, crossing a pair of saddles formed by the interplay between nuclear attraction, interaction with the electric field, and electron–electron repulsion.
Naturally, in the molecular case, the three two-dimensional potentials~\eqref{eq:potA},~\eqref{eq:potB}, and~\eqref{eq:potC} exhibit saddles located at different positions and with different properties, which may affect the ionization signals.

For the helium atom, Eckhardt and Sacha showed that the saddles lie along two straight lines inclined at an angle of $\alpha=\pm\pi/6$ with respect to the $z$-axis, independent of the field strength $F$~\cite{Sacha01,Eckhardt01}.
This allows for a coordinate transformation that restricts the electron motion to these lines~\cite{Sacha06}.
In contrast, for molecules, the simple scaling symmetry of the atomic case is lost, and the saddle positions can no longer be parametrized by straight lines.

\begin{figure}[tb!]
 \includegraphics[scale=1.]{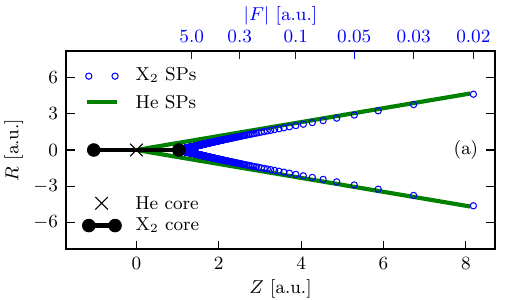}\\
 \includegraphics[scale=1.]{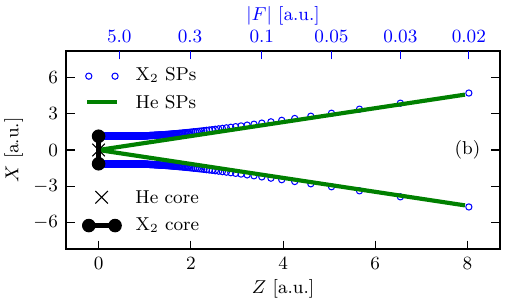}\\
 \includegraphics[scale=1.]{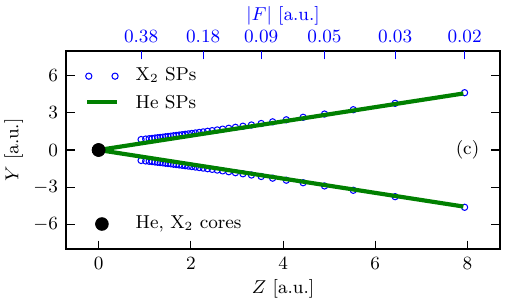}
 \caption{\label{fig:sps}(Color online) The saddle points (SPs) of the model potentials $V_{\parallel}$ (a), $V_{\perp 2}$ (b), and $V_{\perp 3}$ (c) for a diatomic molecule X$_2$ with internuclear distance $d = 2.28$ a.u. are shown in blue. For comparison, saddle points for the helium atom are shown in green. In the case of $V{\perp 3}$, solutions exist only for field amplitudes $|F| < 0.39$ a.u. The upper axis displays the absolute peak field amplitude $|F|$ corresponding to the molecular saddle positions.}
\end{figure}

Figure~\ref{fig:sps} shows the saddle positions (blue circles) for various field strengths and for the three molecular configurations. For comparison, the straight lines corresponding to the helium atom placed at the origin are also shown.
At low field strengths, the saddles are located far from the molecule and approximately follow these straight lines.
However, at higher field strengths, the saddles move closer to the nuclei and deviate from the straight-line paths.
To quantify this deviation, we define
\begin{equation}
 \delta = \frac{|x_s^{\textnormal{Mol}}-x_s^{\textnormal{He}}|}{2\max{\left(|x_s^{\textnormal{Mol}}|,|x_s^{\textnormal{He}}|\right)}},
\end{equation}
where  $x_s^{\textnormal{Mol}}$ is the $R$, $X$ or $Y$ coordinate of the saddle in the molecular case (depending on the configuration), and $x_s^{\textnormal{He}}$ is the corresponding coordinate for helium.
Using internuclear distances for N$_2$, O$_2$ and S$_2$ (see Table~\ref{tab:epsilon}) and peak field amplitudes of $F = 0.22$ a.u. for N$_2$, $F = 0.16$ a.u. for O$_2$, and $F = 0.12$ a.u. for S$_2$, we find that the maximum deviations fall within the range $\delta = (8.6 \pm 4.0)\,\%$.
Thus, the straight-line approximation still provides a reasonable representation of the saddle trajectories, as originally proposed in~\cite{Eckhardt01}.

For the configurations represented by $V_{\parallel}$ (Eq.~\eqref{eq:potA}, Fig.~\ref{fig:config}(a)) and $V_{\perp 2}$ (Eq.~\eqref{eq:potB}, Fig.~\ref{fig:config}(b)), we apply the coordinate transformation $x_i = (-1)^i r_i/2$, $z_i = \sqrt{3} r_i/2$, and $y_i = 0$ for $i = 1,2$~\cite{Sacha06}, in Eqs.~\eqref{interacion} and~\eqref{Mol_pot}.
In the third configuration, represented by $V_{\perp 3}$ (Eq.~\eqref{eq:potC}, Fig.~\ref{fig:config}(c)), the roles of $x_i$ and $y_i$ are interchanged.
The electron–core interaction terms $\widetilde{V}$ for the three configurations are then given by
\begin{eqnarray}
&\widetilde{V}_{\parallel}(r_1,r_2)& = -\sum_{i=1}^2 \frac{1}{\sqrt{r_i^2 + r_i d \sqrt{3}/2 + d^2/4}} \nonumber \\
&&+\frac{1}{\sqrt{r_i^2 - r_i d \sqrt{3}/2 + d^2/4}},\\
&\widetilde{V}_{\perp 2}(r_1,r_2)& = -\sum_{i=1}^2 \frac{1}{\sqrt{r_i^2 + r_i d/2 + d^2/4}}\nonumber\\
&&+\frac{1}{\sqrt{r_i^2 - r_i d/2 + d^2/4}}\\
&\widetilde{V}_{\perp 3}(r_1,r_2)& = -\frac{2}{\sqrt{r_1^2 + d^2/4}} - \frac{2}{\sqrt{r_2^2 + d^2/4}}.
\end{eqnarray}

The electron–electron repulsion and field interaction terms are identical for all three configurations
\begin{eqnarray}
V_{12}(r_1,r_2) &=& \frac{1}{\sqrt{(r_1 - r_2)^2 + r_1 r_2}}, \label{eq:eerep}\\
V_I(r_1,r_2,t) &=& \frac{\sqrt{3}}{2} F(t) (r_1 + r_2). \label{eq:field}
\end{eqnarray}

The resulting Hamiltonian for the restricted-geometry model reads
\begin{equation}
H = \frac{p_1^2 + p_2^2}{2} + \widetilde{V} + V_{12} + V_I, \label{eq:modelhamilt}
\end{equation}
where $\widetilde{V}$ is the only term that distinguishes between the three configurations.

\subsection{Model Assumptions and Limitations}

Since our more specific goal in this work is to explore the predictive power and limitations of the restricted-geometry quantum framework within Eckhard-Sacha approach as applied to NSDI in diatomic molecules, a few comments are in order before we proceed to the numerical implementation and discussion of the results.

First, we note that the starting point of our approach is a hydrogen-like Hamiltonian. At this level, the internal structure of the target (atom or molecule) is not explicitly included, similar to the SAE approximation commonly used in the description of HHG or ATI. Regardless of the atomic species, SAE assumes that only one electron interacts with the field, while the influence of the remaining electrons is incorporated into an effective potential. In our case, we include two active electrons, and the only parameter in the Hamiltonian that reflects the molecular species is the internuclear distance $d$. In this sense, the model describes hydrogen-molecule-like systems. However, in the numerical implementation (see sec.~\ref{sec:ffsimu}), we also introduce a parameter that allows us to set the ionization potential, thereby tailoring the model to a specific molecular species.

In the following, we start with the two-electron ground state function that has spacial part symmetric under particle exchange, i.e. spin part corresponds to a singlet -  $\sigma$ orbital.
An open question remains: to what extent can molecular symmetries, such as the distinction between $\sigma$ and $\pi$ orbitals, be mapped onto the initial wave function within the any restricted-geometry quantum framework? Similar considerations have been made in the context of differences between the ground and excited states of helium~\cite{Eckhardt2008}, as well as in studies of triple ionization~\cite{Thiede2018,Efimov2019,Efimov2020,Prauzner-Bechcicki2021,Efimov2021,Efimov2023}. This issue may serve as a fruitful direction for future research, particularly in the context of extending the model to more complex molecular systems.

Second, the simplified model includes only two orientations of the molecule with respect to the laser polarization axis. This limits the ability to analyze ionization for randomly oriented molecules; averaging over just two orientations is not meaningful. Nevertheless, the two considered orientations, parallel and perpendicular, are experimentally accessible and relevant.

Third, nuclear motion and its influence on ionization channels are neglected in the present approach. At this stage, there appears to be no straightforward way to incorporate this degree of freedom within the current framework.

Fourth, an extension of the approach to triple and higher-order ionization is, in principle, possible. Such generalizations would follow similar lines to those developed for atomic systems~\cite{Thiede2018,Prauzner-Bechcicki2021,Efimov2021}.

Finally, extending the framework beyond diatomic molecules is conceivable. However, in such cases, molecular symmetries are expected to play a much more significant role and would likely require a more elaborate treatment.

\section{Numerical approach\label{sec:ffsimu}}
The first obstacle in the numerical implementation of the proposed model is the presence of singularities in the Coulomb terms. To eliminate these singularities, we introduce a cut-off parameter $\epsilon > 0$ such that
\begin{equation}
\frac{1}{\sqrt{f(r_1,r_2,d)}} \longrightarrow \frac{1}{\sqrt{f(r_1,r_2,d)+\epsilon}}. \end{equation}
This standard smoothing procedure~\cite{Grobe1992,Grobe93,Bauer1997,Lappas1998,Liu1999,Lein00,Ruiz2006,Prauzner-Bechcicki2007,Thiede2018,Efimov2025} allows us to regularize the Coulomb singularities. Both $\epsilon$ and the internuclear distance $d$ can be used to tune the spectrum of the Hamiltonian~(\ref{eq:modelhamilt}), enabling us to approximate the ionization energies of real molecules. In our approach, we take $d$ from experimental data and vary only $\epsilon$ (see Table~\ref{tab:epsilon}).

In the following, we consider three diatomic molecules: N$_2$ ($d = 2.07$~a.u.), O$_2$ ($d = 2.28$~a.u.), and S$_2$ ($d = 3.57$~a.u.). For each potential ($V_{\parallel}$, $V_{\perp 2}$, and $V_{\perp 3}$), an appropriate value of $\epsilon$ is determined numerically by solving the field-free time-dependent Schrödinger equation in imaginary time~\cite{Grobe93}, both for the neutral molecule (using the full Hamiltonian~(\ref{eq:modelhamilt})) and for the corresponding molecular ion (by taking the limit $r_2 \to \infty$ in~(\ref{eq:modelhamilt})). The resulting ionization energies are then compared to experimental values.

Each model molecule is thus characterized by an internuclear distance $d$ and three cut-off parameters $\epsilon$, one for each potential configuration. These values, along with the corresponding ionization energies, are listed in Table~\ref{tab:epsilon}. We find that it is always possible to obtain a good approximation to the experimental ionization energies, although the different subspaces require different values of $\epsilon$.

\begin{table}[t!]
 \begin{center}
  \begin{ruledtabular}
   \begin{tabular}{rrrrr}
   & & $V_{\parallel}$ & $ V_{\perp 2}$ & $ V_{\perp 3}$ \\
   \noalign{\smallskip}
  N$_2$: $d=2.07$~a.u.  & exp. & $\epsilon = 1.6$ & $\epsilon = 1.2$ & $\epsilon = 1.1$ \\
  \hline\noalign{\smallskip}
  $E_g$ [eV]& $-42.7$ & $-42.0$ & $-41.9$ & $-41.1$ \\
  $E_g^+$ [eV]& $-27.1$ & $-28.1$ & $-28.8$ & $-28.7$ \\
  $E_I^+$ [eV]& $15.6$ & $13.9$ & $13.1$ & $12.4$ \\
  \hline\noalign{\smallskip}
  O$_2$: $d=2.28$~a.u. &  exp. & $\epsilon = 2.3$ & $\epsilon = 1.9$ & $\epsilon = 1.6$ \\
  \hline\noalign{\smallskip}
  $E_g$ [eV]& $-36.2$ & $-36.1$ & $-36.0$ & $-36.3$ \\
  $E_g^+$ [eV]& $-24.1$ & $-24.3$ & $-24.6$ & $-25.2$ \\
  $E_I^+$ [eV]& $12.1$ & $11.8$ & $11.4$ & $11.1$ \\
  \hline\noalign{\smallskip}
  S$_2$: $d=3.57$~a.u.  & exp. & $\epsilon = 2.7$ & $\epsilon = 1.3$ & $\epsilon = 1.2$ \\
  \hline\noalign{\smallskip}
  $E_g$ [eV]& $-28.1\phantom{^a}$ & $-29.8$ & $-30.0$ & $-28.4$ \\
  $E_g^+$ [eV]& $-18.7$\footnote[1]{For S$_2$, no experimental value of $E_g^{+}$ could be found in the literature. The value listed in the table was estimated by assuming that the ratio of ground-state to ionic-state ionization energies is the same as for O$_2$, i.e., $E_g^{+}$(S$_2$) was approximated as $E_I^+$(S$_2$) $\times$ $E_g^+$(O$_2$)$/E_I^+$(O$_2$). This approximation is used solely for the purpose of parameter tuning and does not affect the generality of the model.}
& $-20.1$ & $-21.5$ & $-20.9$ \\
  $E_I^+$ [eV]& $9.4\phantom{^a}$ & $9.7$ & $8.5$ & $7.5$ \\
 \end{tabular}
 \end{ruledtabular}
 \end{center}
 \caption{\label{tab:epsilon} Ground state energies of the neutral molecule ($E_g$), ground state energies of the singly ionized molecule ($E_g^+$) and first ionization energies ($E_I^+=E_g^+ - E_g$) for the model versions of N$_2$, O$_2$ and S$_2$, compared to the experimental (exp.) energies of the real molecules~\cite{Herzberg14,Lide96}. The origin of the energy axis is the ground state energy of the doubly ionized molecule, E$_g^{2+} = 0$~eV.}
\end{table}

The time-dependent Schrödinger equation is solved on an $L \times L$ square grid using the split-operator technique combined with the fast Fourier transform (FFT) algorithm~\cite{Feit1982,Dattoli1996}.
The grid size $L$ is first adjusted during the imaginary-time propagation to ensure proper convergence of the ground state. Once the system is exposed to the laser pulse, the grid is extended to accommodate rescattered electrons. Specifically, the grid size is chosen to satisfy the condition $L \gg 2|F|/\omega^2$, i.e., it significantly exceeds the classical quiver distance.
The extension is performed while keeping the grid-point density constant. To prevent spurious reflections of the wave function at the grid boundaries, absorbing boundary conditions are applied.

To calculate the ionization yield, we follow the method used in the atomic case~\cite{Dundas99,Prauzner-Bechcicki2007,Prauzner-Bechcicki2008}. The configuration space is divided into three types of regions:
\begin{itemize}
    \item region $M$, corresponding to the neutral molecule,
    \item regions $S_i$, corresponding to singly ionized molecule,
    \item and regions $D_i$, corresponding to doubly ionized molecule 
\end{itemize}
with $i = 1, \ldots, 4$. This division is based solely on geometric criteria:
\begin{itemize}
    \item if $|r_1| > 8$ a.u. and $|r_2| > 8$ a.u., the wave function is in one of the $D_i$ regions;
    \item if ($|r_1| > 14$ a.u. and $|r_2| < 8$ a.u.) or ($|r_1| < 8$ a.u. and $|r_2| > 14$ a.u.), it is in one of the $S_i$ regions;
    \item otherwise, it is in the $M$ region.
\end{itemize}
The ionization yield is then calculated as the probability flux between these regions. The detailed procedure is analogous to that described in~\cite{Prauzner-Bechcicki2008}.

To obtain momentum distributions, we follow the method proposed by Lein \textit{et al.}~\cite{Lein00} and later applied to studies of double~\cite{Prauzner-Bechcicki2008} and triple ionization in atoms~\cite{Efimov2021}. The whole configuration space is divided into four regions:
\begin{itemize}
    \item $M_{\text{in}}$ ($|r_1|, |r_2| < a$),
    \item $M_1$ ($|r_1| \ge a$, $|r_2| < a$),
    \item $M_2$ ($|r_1| < a$, $|r_2| \ge a$),
    \item and $M_{\text{out}}$ ($|r_1|, |r_2| \ge a$).
\end{itemize}
The parameter $a$ is chosen to exceed the classical quiver distance, ensuring that once an electron reaches this distance, it is unlikely to return. In region $M_{\text{out}}$, Coulomb interactions are neglected, and the wave function is evolved in momentum space via multiplication by a phase factor. In regions $M_1$ and $M_2$, the Coulomb interaction between electrons and the interaction with the nuclei for the more distant electron are neglected. In $M_{\text{in}}$, the full Hamiltonian is used.

Wave function transfer between regions is performed using the “smooth cutting and coherent adding” procedure~\cite{Lein00}. The two-electron momentum distribution is obtained by coherently collecting the wave function from all regions, while smoothly removing contributions corresponding to the neutral and singly ionized molecule. The modulus square of the resulting wave function in momentum space yields the two-electron momentum distribution. For a comprehensive description of the wave function transfer technique and the construction of two-electron momentum distributions, we refer the reader to~\cite{Prauzner-Bechcicki2008}.

The ion momentum distribution is then obtained as the marginal distribution of the variable
\begin{equation}
p_{\parallel}^{2+} = -\frac{\sqrt{3}}{2}(p_1 + p_2).
\end{equation}

Before moving on, we add one final comment.
Both the actual molecules and the models described above differ in their ionization energies. A meaningful comparison between different cases can be based on an approximate classical scaling. Consider two molecules with ground-state energies $E$ and $E'$. To relate the two, we introduce an effective quantum number $q := \sqrt{E'/E}$ and consider the scaling of the classical Hamiltonian
\begin{equation}
 \begin{split}
 H(q^{-2}r_{i},qp_i,q^{-3}t,q^4F_0,q^3\omega,\phi,q^{-4}\epsilon,q^{-2}d) =\\
 q^2H(r_i,p_i,t,F_0,\omega,\phi,\epsilon,d)\label{eq:resc}
 \end{split}
\end{equation}
This implies that the field parameters scale as 
\begin{eqnarray} F_0' &=& F_0\, q^4, \label{eq:fresc}\\
\omega' &=& \omega\, q^3. \label{eq:wresc}
\end{eqnarray}

The usefulness of this classical scaling lies in its ability to compensate for the differences in ionization energies between the model potentials. Since the ground-state energy sets the characteristic scale of electron motion, molecules with different $E_g$ respond most similarly when the driving field parameters $(F_0,\omega)$ are rescaled according to Eqs.~(\ref{eq:fresc})--(\ref{eq:wresc}). In practice, this scaling aligns the effective barrier heights and excursion amplitudes experienced by the electrons, allowing ionization yields of systems with different binding energies to be compared on an approximately equal footing. Although the procedure is not exact, since it combines quantum-mechanical energies with a classical scaling, it captures the dominant trends and substantially improves the interpretability of cross-molecular ionization-yield comparisons.

The scaling law~(\ref{eq:resc}) also suggests that the cut-off parameter $\epsilon$ and the internuclear distance $d$ can be mapped accordingly. While perfect agreement cannot be expected, since the ground-state energies are obtained from quantum mechanical calculations, the trend is consistent. For example, rescaling the O$_2$ parameters to match N$_2$ yields $d = 1.96$~a.u. and $\epsilon = 1.7$ ($V_{\parallel}$), $\epsilon = 1.4$ ($V_{\perp 2}$), and $\epsilon = 1.2$ ($V_{\perp 3}$), which compare reasonably well with the actual N$_2$ values listed in Table~\ref{tab:epsilon}. In fact, $d$ is slightly smaller and $\epsilon$ slightly larger than the empirically adjusted values, so the deviations partially compensate. For S$_2$, however, the rescaled values deviate more significantly. Thus, while the scaling is not perfectly implemented, the discrepancies are comparable to those observed in the saddle-point positions.

In the next section we take advantage of the classical scaling of the field parameters to improve the comparability of ionization yields for the three molecular species N$_2$, O$_2$, and S$_2$.

\begin{table}[t!]
 \begin{center}
  \begin{ruledtabular}
   \begin{tabular}{cccc}
    Molecule & $E_g$ [eV]& $\omega$ [a.u.] & $F_{0}$ [a.u.]\\
    \hline
    N$_2$ & $-42.0$ & 0.075 &  0.22\\
    O$_2$ & $-36.1$ & 0.060 &  0.16\\
    S$_2$ & $-29.8$ & 0.045 &  0.11\\
   \end{tabular}\caption{Frequencies $\omega$ and example peak field amplitudes $F_{0}$ in a.u. that coincide with respect to classical scaling. We used the listed ground state energies $E_g$ from potential $V_{\parallel}$ (see Table~\ref{tab:epsilon}) and selected the O$_2$ frequency and peak field amplitude as reference values for rescaling to obtain the respective values for N$_2$ and S$_2$ listed above.\label{tab:resc}}
   \end{ruledtabular}
 \end{center}
\end{table}

\section{Results}
In this section, we discuss the results of propagating the ground-state wave functions for the three molecules N$_2$, O$_2$, and S$_2$, using the three model potentials in the presence of a linearly polarized laser field derived in Sec.~\ref{model}.
Results for the potential $V_{\perp 3}$ are not shown, as they closely resemble those obtained for the other two configurations and do not reveal any additional features beyond what is already evident from the comparison between $V_{\perp 2}$ and $V_{\parallel}$.
Unless stated otherwise, all simulations are performed using the same laser pulse as in~\cite{Eckhardt2010}, consisting of $n_c = 5$ field cycles and a carrier-envelope phase $\phi = 0$.

In the following subsections, we compare ionization yield curves for different molecular alignments and species, and present results for momentum distributions. We emphasize once again that the analysis of molecular alignment is limited to two extreme orientations, parallel and perpendicular, while the molecular species differ only in their internuclear distance and ionization energy.

\subsection{Ionization Yields\label{sec:ysimu}}

\begin{figure}[!t]
 \centering
  \includegraphics[scale=1]{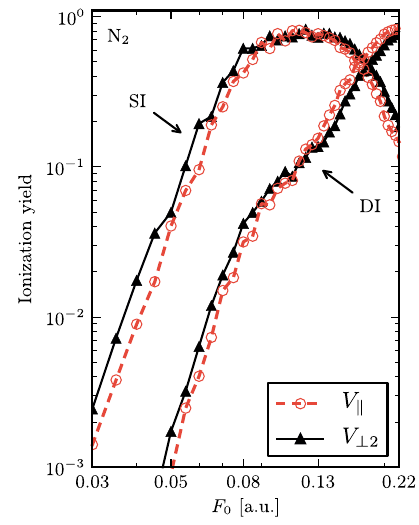}
 \caption{\label{fig:compN2}(Color online) Single ionization (SI) and double ionization (DI) yield curves of potentials $V_{\parallel}$ and $V_{\perp 2}$ at a frequency of $\omega = 0.075$ a.u. for N$_2$.}
\end{figure}
\begin{figure}[!t]
 \centering
  \includegraphics[scale=1]{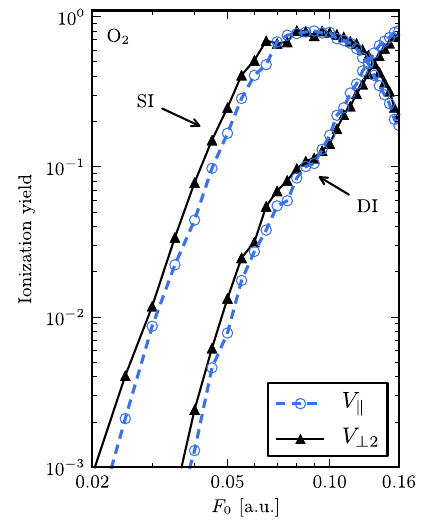}\\
 \caption{\label{fig:compO2}(Color online) Single ionization (SI) and double ionization (DI) yield curves of potentials $V_{\parallel}$ and $V_{\perp 2}$ at a frequency of $\omega = 0.06$ a.u. for O$_2$. }
\end{figure}
\begin{figure}[!t]
 \centering
  \includegraphics[scale=1]{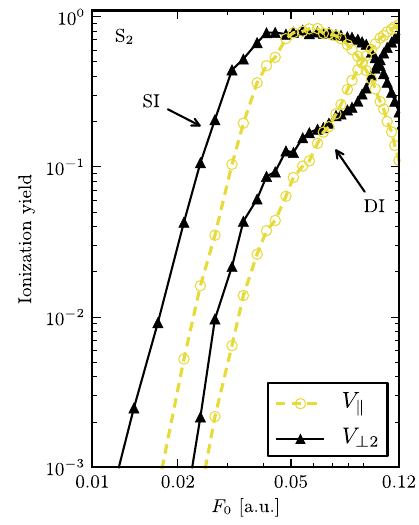}
 \caption{\label{fig:compS2}(Color online) Single ionization (SI) and double ionization (DI) yield curves of potentials $V_{\parallel}$ and $V_{\perp 2}$ at a frequency of $\omega = 0.045$ a.u. for S$_2$.}
\end{figure}

We start by showing the ionization yield curves for the three model molecules in Figs.~\ref{fig:compN2}–\ref{fig:compS2}. Each figure presents both single ionization (SI) and double ionization (DI) yields. Circles correspond to the parallel configuration ($V_{\parallel}$), while filled triangles represent the perpendicular configuration ($V_{\perp 2}$). The choice of different frequencies $\omega$ is motivated by classical scaling (see Table~\ref{tab:resc}).

The characteristic knee structure in the DI yield curve is visible in all three cases, indicating the presence of NSDI channel. In each case, the position of the knee correlates well with the saturation of the SI yield, as expected. The observed shift of the knee onset towards lower field amplitudes when moving from N$_2$ to O$_2$ and then to S$_2$ reflects the decreasing ionization potential across this sequence. This trend is not attributed to the change in frequency~\cite{Thiede2024}.

When comparing the yield curves, the different configurations do not lead to visible differences in the case of N$_2$ and O$_2$ (see Figs.~\ref{fig:compN2} and~\ref{fig:compO2}), whereas for S$_2$ (Fig.~\ref{fig:compS2}) the difference is clearly pronounced.
Several factors may contribute to the observed discrepancy between N$_2$ (or O$_2$) and S$_2$. One possible explanation lies in the actual differences in ionization energies between the two configurations for the same molecular species. For S$_2$, these differences are $\Delta E_I^+ = 1.2$ eV and $\Delta E_I^{2+} = 1.4$ eV. However, for N$_2$, the differences are smaller ($\Delta E_I^+ = 0.8$ eV and $\Delta E_I^{2+} = 0.7$ eV) and the ionization yields for different configurations are virtually identical (compare Figs.~\ref{fig:compN2} and~\ref{fig:compS2}).

Another contributing factor is the difference in the cut-off parameter $\epsilon$ between the two alignments. For S$_2$, the difference is relatively large ($\Delta\epsilon = 1.4$), while for O$_2$ and N$_2$ it is smaller ($\Delta\epsilon = 0.4$ in both cases). In all three molecules, $\epsilon$ is larger for the parallel configuration than for the perpendicular one. As a result, the electron–electron repulsion term in $V_{\parallel}$ is shallower than in $V_{\perp 2}$, leading to a reduction in electron–electron interaction. Since NSDI is rooted in electron–electron correlations, it strongly depends on this interaction term. Therefore, one can expect the DI yield for $V_{\parallel}$ to be lower than for $V_{\perp 2}$ in the NSDI regime. This effect should be particularly pronounced for S$_2$, and indeed, Fig.~\ref{fig:compS2} shows a smaller DI yield for $V_{\parallel}$. We thus consider this model-specific effect to be the most likely explanation for the behavior of the S$_2$ ionization yields.

Overall, the results indicate that ionization yields vary little with molecular orientation. This is consistent with the findings of Prauzner-Bechcicki~\textit{et al.} in the classical model analysis~\cite{Prauzner-Bechcicki2005}. However, experimental results by Zeidler~\textit{et al.}~\cite{Zeidler05} show that $30\,\%$ more N$_2^{2+}$ ions and $19\,\%$ more N$_2^+$ ions are produced for parallel alignment compared to perpendicular alignment. Moreover, $S$-matrix calculations suggest that the electronic structure of the molecular orbital influences ionization~\cite{Faria08,Jia09}. This indicates that the symmetry of the molecular orbital relative to the laser polarization axis may play a significant role. While this observation could explain the discrepancy between experimental data and our numerical results, a full test of this hypothesis is, as noted earlier, beyond the predictive power of the present model.

\begin{figure}[!tb]
 \centering
  \includegraphics[scale=1]{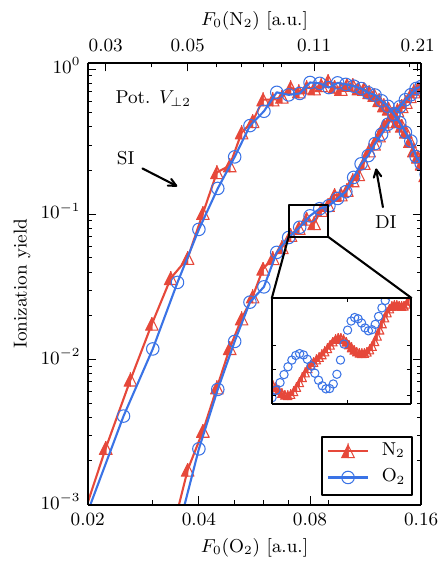}
 \caption{\label{fig:O2N2_Bmid}(Color online) Comparison of the single ionization (SI) and double ionization (DI) yields of potential $V_{\perp 2}$ of N$_2$ ($\omega = 0.075$ a.u.) and O$_2$ ($\omega = 0.06$ a.u.). The peak field amplitudes $F_0$ for N$_2$ (top axis) have been rescaled according to Eq.~\eqref{eq:fresc}.}
\end{figure}
\begin{figure}[!tb]
 \centering
  \includegraphics[scale=1]{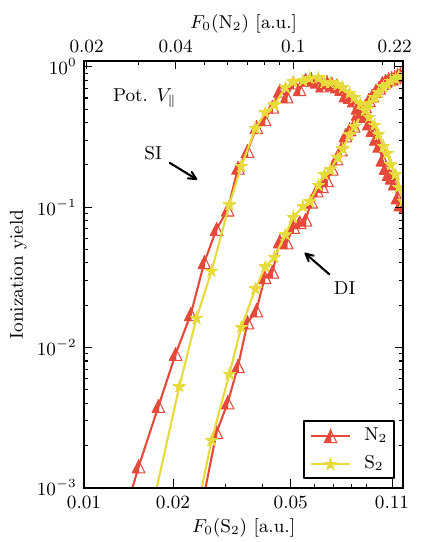}
 \caption{\label{fig:N2S2_Amid}(Color online) Comparison of the single ionization (SI) and double ionization (DI) yields of potential $V_{\parallel}$ of N$_2$ ($\omega = 0.075$ a.u.) and S$_2$ ($\omega = 0.045$ a.u.). The peak field amplitudes $F_0$ for N$_2$ (top axis) have been rescaled according to Eq.~\eqref{eq:fresc}.}
\end{figure}
\begin{figure}[!tb]
 \centering
  \includegraphics[scale=1]{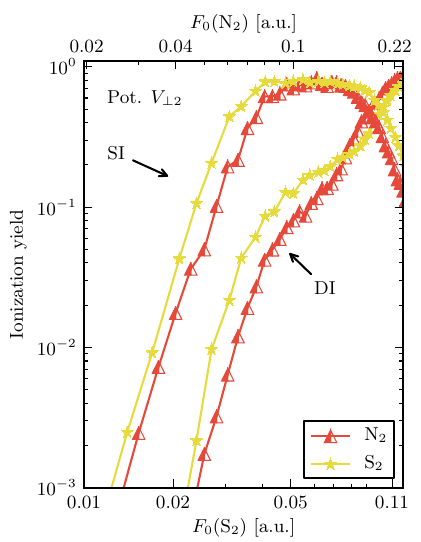}
 \caption{\label{fig:N2S2_Bmid}(Color online) Comparison of the single ionization (SI) and double ionization (DI) yields of potential $V_{\perp 2}$ of N$_2$ ($\omega = 0.075$ a.u.) and S$_2$ ($\omega = 0.045$ a.u.). The peak field amplitudes $F_0$ for N$_2$ (top axis) have been rescaled according to Eq.~\eqref{eq:fresc}.}
\end{figure}

To further compare the studied molecular species, we apply the classical scaling described in the previous section. As an example, we first show a comparison of ionization yields for the perpendicular orientation ($V_{\perp 2}$) between N$_2$ and O$_2$ in Fig.~\ref{fig:O2N2_Bmid}, and for the parallel orientation ($V_{\parallel}$) between N$_2$ and S$_2$ in Fig.~\ref{fig:N2S2_Amid}. In both cases, a remarkably good agreement in the global trend is observed, highlighting once again the model’s limited ability to distinguish between molecular species (i.e., the molecular species are differentiated only by a change in the internuclear distance $d$, not by their ground‑state symmetry).

A similar behavior is found for all other molecule pairs with the same orientation, except for those involving the $V_{\perp 2}$ potential of S$_2$. For instance, Fig.~\ref{fig:N2S2_Bmid} illustrates the differences in this case for N$_2$ and S$_2$. Again, we attribute the discrepancy to the strong reduction of electron–electron repulsion caused by the choice of the cut-off parameter $\epsilon$.

Eremina \textit{et al.} argued in~\cite{Eremina04} that SI depends on the molecular structure. Since SI is the first step in the rescattering mechanism~\cite{Corkum93,Kulander1993}, they also expected the molecular symmetry to influence the NSDI process. Indeed, their measurements confirmed such a dependence~\cite{Eremina04}. Similarly, Guo \textit{et al.} reported a significantly reduced DI rate for O$_2$ and attributed it to the molecular structure of the molecule~\cite{Guo98}.

While we again reach the limits of the predictive power of the present model, one experimental suggestion may be formulated: the field amplitudes and frequencies should be adapted according to classical scaling. If differences in ionization yields persist under such conditions, this would indicate that the molecular structure indeed plays a significant role in the DI process.

A detailed analysis of the DI yield curves reveals the presence of small-scale, non-monotonic variations, including local extrema. These features were initially identified during a coarse-grained scan along the field amplitude axis and are highlighted in the inset of Fig.~\ref{fig:O2N2_Bmid}. The inset presents a segment of the yield curves obtained with finer steps in $F_0$, ruling out the possibility that the observed extrema are numerical artifacts. Similar structures are visible in Fig.~\ref{fig:compN2} for $V_{\parallel}$ near $F_0 \approx 0.09$ a.u. and for $V_{\perp 2}$ near $F_0 \approx 0.11$ a.u., as well as in Fig.~\ref{fig:compS2} for $V_{\perp 2}$ near $F_0 \approx 0.05$ a.u. Increasing the resolution of the field amplitude scan by reducing the step size $\Delta F_0$ reveals even more of these genuine dynamical features.
These structures are not restricted to a particular molecule or potential and can be observed across different frequencies, albeit at different values of $F_0$. Furthermore, extrema visible in the DI yield for $V_{\parallel}$ often have counterparts at similar $F_0$ values in the DI yield for $V_{\perp 2}$.

Panfili \textit{et al.}~\cite{Panfili03} identified similar local extrema in the ionization yield of a model helium atom and attributed them to multiphoton resonances between the ground state and low-lying bound states. In their simulations, they used a sine pulse with a trapezoidal envelope and found that increasing the number of field cycles $n_c$ enhanced the amplitude of the maxima.

\begin{figure}[!t]
 \centering
  \includegraphics[scale=1]{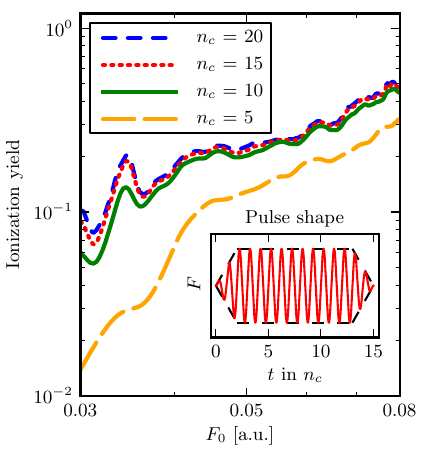}
 \caption{\label{fig:S2potBTn}(Color online) Resonances for different numbers of field cycles $n_c$ in the knee region of S$_2$ potential $V_{\perp 2}$ at $\omega = 0.06$ a.u. An example for $n_c = 15$ of the used laser pulse with trapezoidal envelope and two cycle turn-on time is displayed as an inset.}
\end{figure}

To test this behavior, we repeated our simulations using a trapezoidal pulse shape and verified that the extrema become more pronounced with increasing pulse duration. This is illustrated in Fig.~\ref{fig:S2potBTn} for the case of S$_2$ and potential $V_{\perp 2}$. 
For longer pulses, the peaks can be associated with resonances among eigenstates of the periodically driven system, i.e., its Floquet states.
Following the approach of Panfili{\it et al.}~\cite{Panfili03}, we estimated the positions of multiphoton resonances in our model by including both the dynamical Stark shift and the ponderomotive shift of the bound-state energies. These estimates, as applied to results for S$_2$ shown in Fig.~\ref{fig:S2potBTn}, indicate that 3-, 4-, and 5-photon resonances between the ground state and the first or second excited state should occur within the same range of field strengths where the local extrema in the double-ionization yield are observed. In this picture, the resonances enhance population transfer into Stark- and field-dressed excited states that act as intermediate channels for double ionization, thereby producing the characteristic peaks.
Similar phenomena have been observed in microwave ionization of atomic systems~\cite{Blumel87,Koch95}, where they were directly linked to Floquet resonances. Although we do not explicitly construct Floquet states in the present model, the stabilization of the ionization yield with increasing pulse duration supports the interpretation in terms of resonance effects.

\subsection{Momentum distributions}
\begin{figure}[!t]
 \centering
  \includegraphics[width=0.5\textwidth]{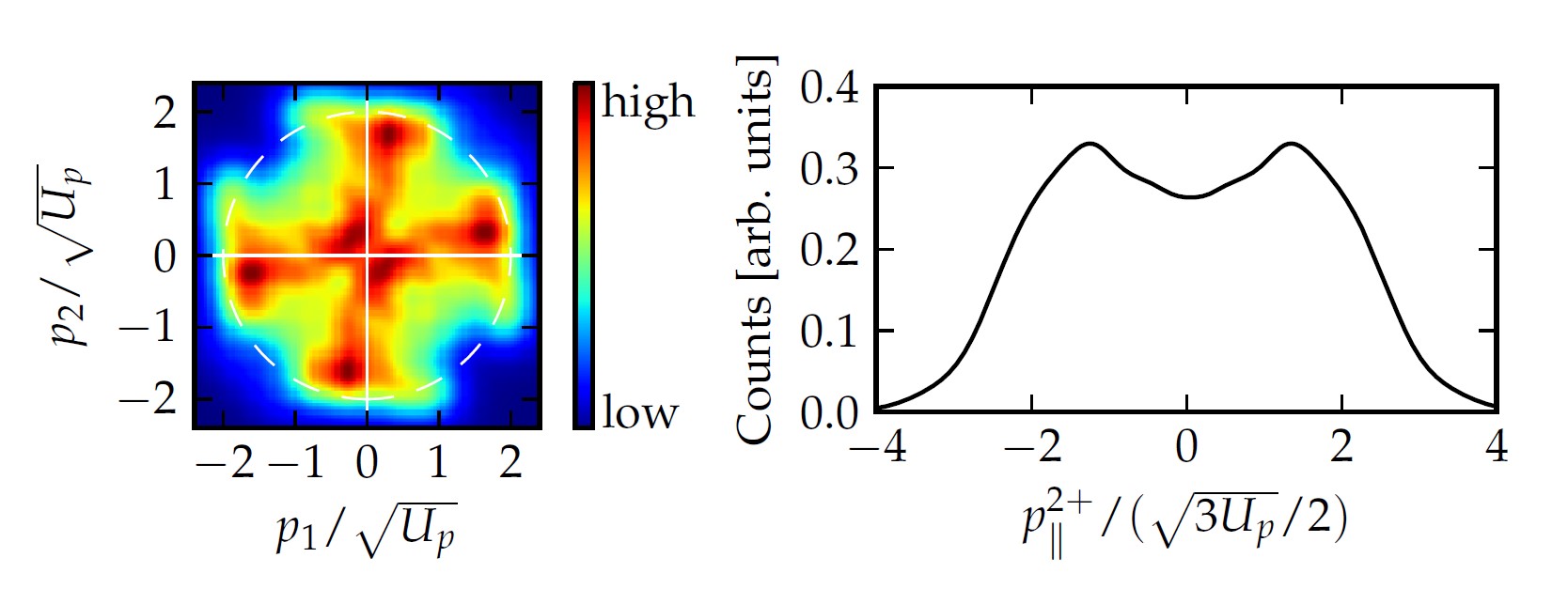}
 \caption{\label{fig:momentum_dist}(Color online) Two-electron momentum distribution (left) and ion momentum distribution (right) for O$_2$ in the parallel configuration, at $F_0 = 0.06$ a.u. and $\omega = 0.06$ a.u., averaged over eight carrier-envelope phases. The dashed circle indicates the maximum momentum that electrons can classically gain in the field.}
\end{figure}

Having identified the limitations of the analyzed model, we decided not to study the dependence of momentum distributions on molecular orientation or species. However, in Fig.~\ref{fig:momentum_dist}, we present both the two-electron momentum distribution and the ion momentum distribution for O$_2$ in the parallel configuration, at $F_0 = 0.06$ a.u. and $\omega = 0.06$ a.u., averaged over eight carrier-envelope phases. These distributions correspond to experimentally accessible observables: two-electron momentum distributions and ion momentum distributions measured along the laser polarization axis.

The obtained results, although averaged over only eight phases, closely resemble those reported by Kübel \textit{et al.} in~\cite{Kubel2013}. In their work, Kübel \textit{et al.} compared the N$_2$ molecule with its atomic counterpart, the Ar atom, and discussed similarities and differences in the two-electron momentum distributions. Both species exhibit asymmetric energy sharing between the escaping electrons, which is associated with the RESI mechanism. In the case of N$_2$, the distributions also show a contribution from low-energy anticorrelated electrons, which leads to a flat-top structure in the ion momentum distribution. This contribution was attributed to the presence of excited states in N$_2^+$~\cite{Kubel2013}.

A similar effect is visible in both the two-electron and ion momentum distributions obtained with our model (see Fig.~\ref{fig:momentum_dist}). We chose to present results for O$_2$ to emphasize the model’s inherent limitation in distinguishing between different molecular species. Despite this limitation, it is noteworthy that the obtained distributions reproduce the experimental results remarkably well. 

There is a difference in the symmetry of the valence orbital between N$_2$ and O$_2$: N$_2$ has a $\sigma_g$ bonding symmetry, while O$_2$ has a $\pi_g$ antibonding symmetry. The acquired results suggest that the introduced model is well suited to study systems with $\sigma$-type symmetry, such as N$_2$, whereas its application to systems with other orbital symmetries may require further extension. Exploring how to incorporate the molecular orbital symmetry degree of freedom into the restricted-geometry framework could be a promising direction for future research.

\section{Conclusions}

We introduced a (1+1)-dimensional quantum model capable of describing non-sequential double ionization (NSDI) in homonuclear diatomic molecules. The model is a straightforward extension of the restricted-geometry Eckhardt--Sacha approach previously used for atoms. We thoroughly discussed the rationale behind this framework and clearly stated its assumptions and limitations. Wherever possible, we pointed out conceivable extensions of the approach. The numerical methodology was also outlined.

Within the proposed model, we analyzed ionization yield curves for two extreme molecular orientations and three different species. The results confirmed the model’s limitations, while simultaneously demonstrating its ability to capture subtle dynamical effects, such as the resonances observed in the DI yield curves. By applying classical scaling, we were able to map the yield curves of different molecules onto each other. When such scaling is applied to experimental data, any remaining differences in the yields may indicate quantum effects, resonances being one such example observed in our simulations.

Finally, the comparison between the obtained two-electron and ion momentum distributions and existing experimental data shows that, despite its simplifying assumptions and clear limitations, the restricted-geometry quantum model is suitable for studying NSDI in diatomic molecules with $\sigma$-type symmetry.

To summarize the physical content of the model, we emphasize which experimental observations it can and cannot reproduce. On the positive side, the model correctly captures (i) the existence and approximate position of the knee structure in the DI yield, (ii) the global dependence of SI and DI yields on the field strength, (iii) resonance-induced non‑monotonic features such as the local extrema traced to multiphoton and Floquet-type resonances, and (iv) characteristic signatures in two-electron and ion momentum distributions, including asymmetric energy sharing and flat-top ion momentum profiles, consistent with experimental observations. However, the model cannot reproduce (i) the experimentally observed dependence of SI and DI yields on molecular orientation, (ii) the suppression of DI in molecules with $\pi$-type valence orbitals, such as O$_2$, (iii) effects associated with multielectron orbital symmetry and configuration mixing, and (iv) orientation averaging or nuclear-motion–induced modifications of ionization channels. These limitations stem directly from the restricted geometry, the absence of explicit orbital symmetry, and the hydrogenic character of the underlying effective potential. A clear identification of which observables fall within the model’s predictive scope helps guide the interpretation of agreements and deviations with experimental data.

\begin{acknowledgments}
The L.C.B., J.H.T and J.S.P.-B. acknowledge the late Bruno Eckhardt with
whom we had the honor of working. The above work was
initiated and stimulated by him.

D.K.E. and M.O. gratefully acknowledge Polish high-performance computing infrastructure PLGrid (HPC Center: ACK Cyfronet AGH) for providing computer facilities and support within computational grant no. PLG/2025/018418.
\end{acknowledgments}


%

\end{document}